\documentclass[twocolumn,3p,times]{elsarticle}

\usepackage{amssymb}

\usepackage[figuresright]{rotating}

\begin{document}

\begin{frontmatter}




\title{Determining the $^{6}$Li Doped Side of a Glass Scintillator for
  Ultra Cold Neutrons}

\author[UWinnipeg]{Blair Jamieson}
\ead{bl.jamieson@uwinnipeg.ca}
\author[UWinnipeg]{Lori Rebenitsch}

\address[UWinnipeg]{
University of Winnipeg\\
515 Portage Avenue\\
Winnipeg, Manitoba}

\begin{abstract}

Ultracold neutron (UCN) detectors using two visually very similar, to
the microscopic level, pieces of optically contacted cerium doped
lithium glasses have been proposed for high rate UCN experiments.  The
chief difference between the two glass scintillators is that one side
is $^6$Li depleted and the other side $^6$Li doped.  This note
outlines a method to determine which side of the glass stack is doped
with $^{6}$Li using AmBe and $^{252}$Cf neutron sources, and a Si
surface barrier detector.  The method sees an excess of events around
the $\alpha$ and triton energies of neutron capture on $^{6}$Li when
the enriched side is facing the Si surface barrier detector.

\end{abstract}


\end{frontmatter}


\section{Introduction}\label{sec:intro}

Several neutron Electric Dipole Moment experiments using UltraCold
Neutrons (UCN) are being planned worldwide \cite{PSI, SNBalashov,
  Gatchina, APSerebov, KKirch, CABaker, YMasuda, IAltarev, RGolub,
  SNS}.  All planned experiments anticipate large increases in the
rate of UCN detected (\textgreater MHz).  New technologies are being
pursued in order to handle the projected increase in rate, while
keeping the efficiency high and the sensitivity to backgrounds arising
from gamma rays and thermal neutrons low.

The benchmark technology in this case is the $^3$He multi-wire
proportional counter (MWPC) \cite{Morris}.  By arranging the density of
the $^3$He to be small, UCN may enter deep into the detector volume
prior to capturing on the $^3$He.  In this way, the reaction products
(deuteron and triton) deposit their energy fully in the surrounding
gas volume, which contains a larger partial pressure of gasses
suitable for creating the avalanche condition necessary for MWPC
operation.  In this way, such detectors have near 100\% efficiency,
and are relatively insensitive to backgrounds (owing to the thinness
of the gas layer).  Such detectors are often limited in their rate
capability, however.  While this can be overcome by segmentation,
other groups have proposed using detection of scintillation light,
which has an inherently faster recovery time than gas counters.

\begin{figure}[htpb]
\begin{center}
\includegraphics[width=0.39\textwidth]{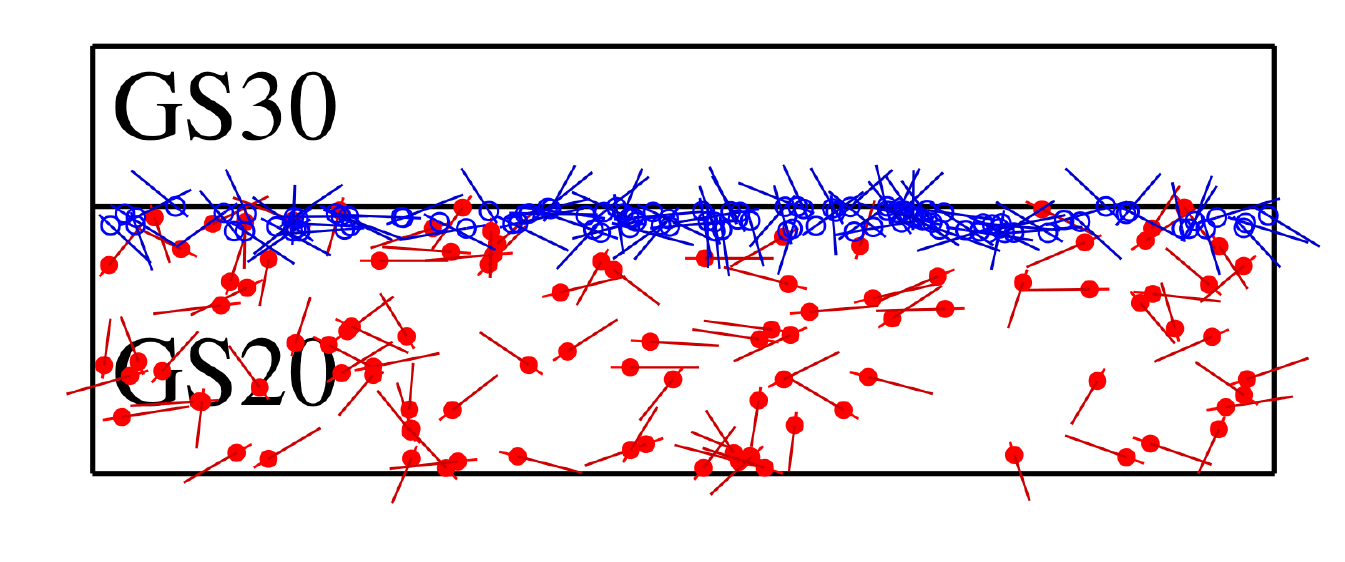}
\caption{\label{fig:capture} Visual example of neutrons capturing in
  the Li-glass stack.  The open circles represent the capture location
  of a UCN incident from the top of the page, and the filled circles
  represent the capture location for thermal neutrons.  The the triton
  and alpha are represented by the long and short thin lines
  respectively, emanating from the capture locations.  In this figure
  four of the triton, and two of the alpha particles escape from the
  bottom of the GS20 layer and could be detected.}
\end{center}
\end{figure}

One technology being pursued involves the use of $^6$Li-doped and
$^6$Li-depleted scintillating glass \cite{Ban, LeFort}.  Such glass is
available from Applied Scintillation Technologies in the UK
\cite{AST-LiE, AST-LiD} as GS30 ($^6$Li-depleted) and GS20
($^6$Li-enriched).  The properties of the glasses are summarized in
Table~\ref{tab:LiGlass}.  The function of a detector based on this
principle is displayed in FIG.~\ref{fig:capture}.  UCN pass through a
$^6$Li-depleted glass layer and then capture on a $^6$Li-enriched
layer.  The neutron capture interaction length in the enriched layer
is $\sim 1.5$~$\mu$m.  Thermal neutrons have a longer interaction
length in the glass, and capture uniformly in the volume of the GS20.
When neutrons capture on $^{6}$Li nuclei, they produce an $\alpha$ and
a triton via the reaction:

\begin{equation}\label{eq:nGlobalLireact}
^6Li+n\rightarrow \alpha\ \textrm{(2.05\ MeV)} + t\ \textrm{(2.73\ MeV)}.
\end{equation}

The mean range of the $\alpha$ is 5.3~$\mu$m, and the mean range of
the triton is 34.7~$\mu$m.  Thinning the glass to thicker than this
allows the full energy of the neutron capture to be deposited in the
glass.  The glass is kept as thin as possible to reduce backgrounds to
the UCN counting.  Keeping the glass thin reduces its efficiency for
seeing thermal neutrons, and minimizing the light output from gamma
rays.

A key enabling technology is the optical contacting of the two glass
layers, which prevents any losses of the reaction products in glue.
Such detectors have demonstrated efficiency rivalling $^3$He counters
\cite{LeFort}.  The decay time of the atomic states in the
scintillating glass is such that most scintillation light is gone
after 50 to 70 ns.  To partially address backgrounds, pulse-shape
discrimination may be used to separate Cherenkov light arising from
gamma rays in the lightguide of this detector.

\begin{table}[hb]
\begin{center}
\begin{tabular}{ | c | c | c | c |}
  \hline
  Scintillator              & GS20                 & GS30                  \\
                            & $^6$Li enriched      & $^6$Li depleted        \\
  \hline
  $^6$Li fraction (\%)      & 95                   & 0.01                   \\
  \hline
  $^6$Li density (cm$^{-3}$) & 2.2$\times$10$^{22}$  &   2.4$\times$10$^{18}$  \\
  \hline
\end{tabular}
\end{center}
\caption{Properties of the glass scintillators used in the detector
  described in this note.}\label{tab:LiGlass}
\end{table}

The glass stack used in this detector is thinned to 60~$\mu$m on the
$^6$Li depleted side and 100~$\mu$m on the $^6$Li doped side.  Making
the GS30 layer as thin as possible reduces the barrier the UCN see
before reaching the GS20 layer, and keeping the GS20 layer thin
reduces the efficiency for capturing thermal neutrons, which are a
background when the main aim of the detector is to detect UCN.

\section{Experimental Method}\label{sec:method}

A 37~MBq AmBe neutron source, and a 185~kBq $^{252}$Cf neutron source
were used to produce thermal neutrons, by leaving them in their lead
and plastic holders respectively.  Two sources were used to produce a
higher rate of neutron captures with the sources available in our lab.
The neutron sources were surrounded with wax blocks to further
thermalize the neutrons.  Thermal neutrons have a long enough
interaction length to capture anywhere within the $^{6}$Li layer of
the scintillating glass, producing an $\alpha$ particle with an energy
of 2.05 MeV and the triton has an energy of 2.73 MeV.  When these
captures happen close enough to the GS20 surface, the $\alpha$ or
triton can be detected with a Si surface barrier detector.  These ions
should only be seen if the GS20 side faces the surface barrier
detector.  In our setup, we use an Ortec A-016-025-500 partially
depleted silicon surface barrier detector biased with a +140~V supply,
to produce an 0.5~$\mu$m thick depletion region to detect energy
deposit when the $\alpha$ or triton reach the detector.

\begin{figure}[htpb]
\begin{center}
\includegraphics[width=0.49\textwidth]{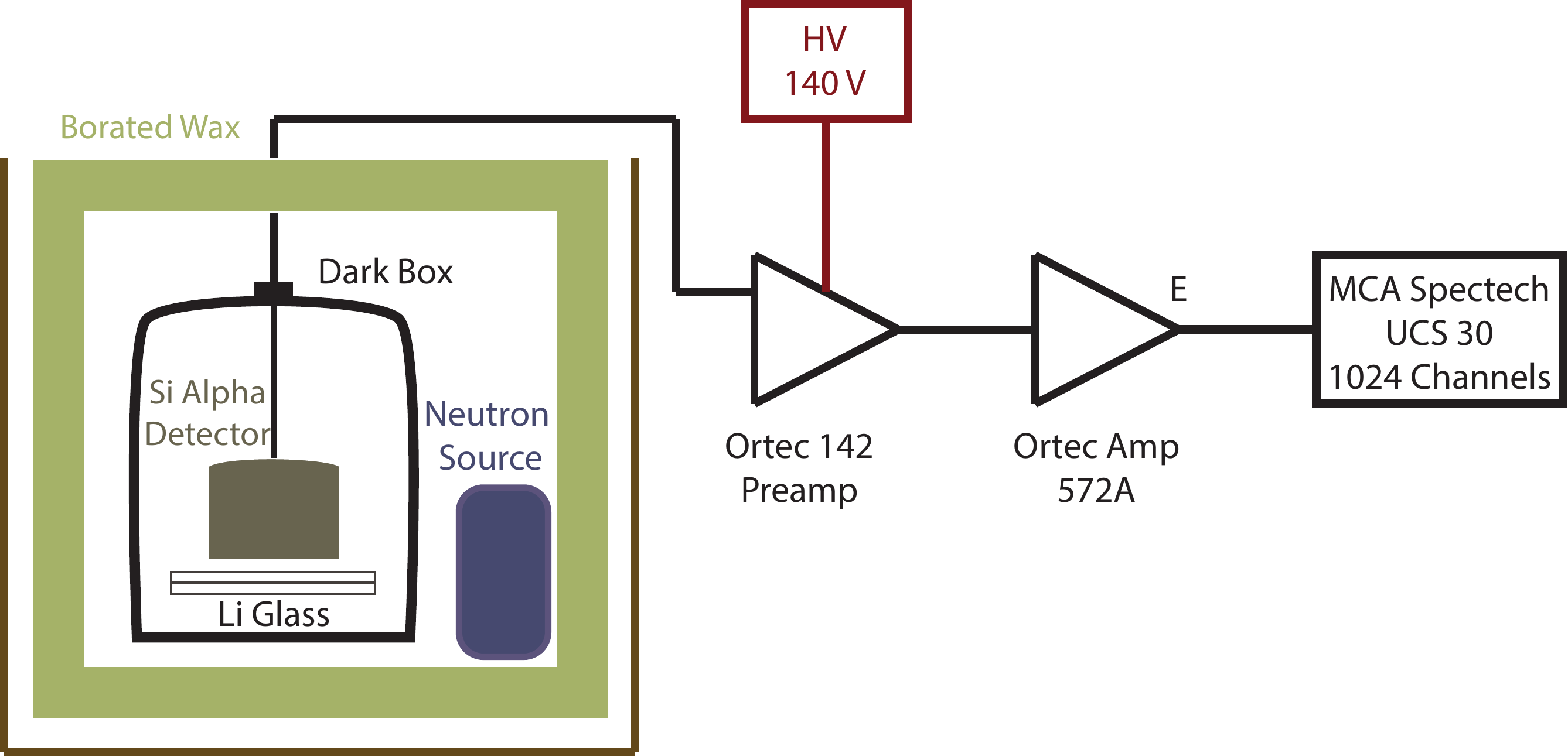}
\caption{\label{fig:setup} Diagram of the experimental setup.}
\end{center}
\end{figure}

The surface barrier detector was placed inside a metal bell which
served as a dark box, and the scintillating glass was placed directly
below the detector in order to minimize the distance the resultant
particles had to travel through air to reach the detector ($\sim
1$~mm).  The entire setup, shown in FIG.~\ref{fig:setup}.

The signal from the surface barrier detector was sent to an Ortec 142A
pre-amplifier, whose output was amplified by an Ortec 572A
spectroscopy amplifier.  The amplified output was sent to a 1024
channel SpecTech Multi-Channel Analyzer to produce spectra of the
energy deposited.

\subsection{Energy Calibration}\label{sec:ecal}

\begin{figure*}[htpb]
\begin{center}
\includegraphics[width=1.0\textwidth]{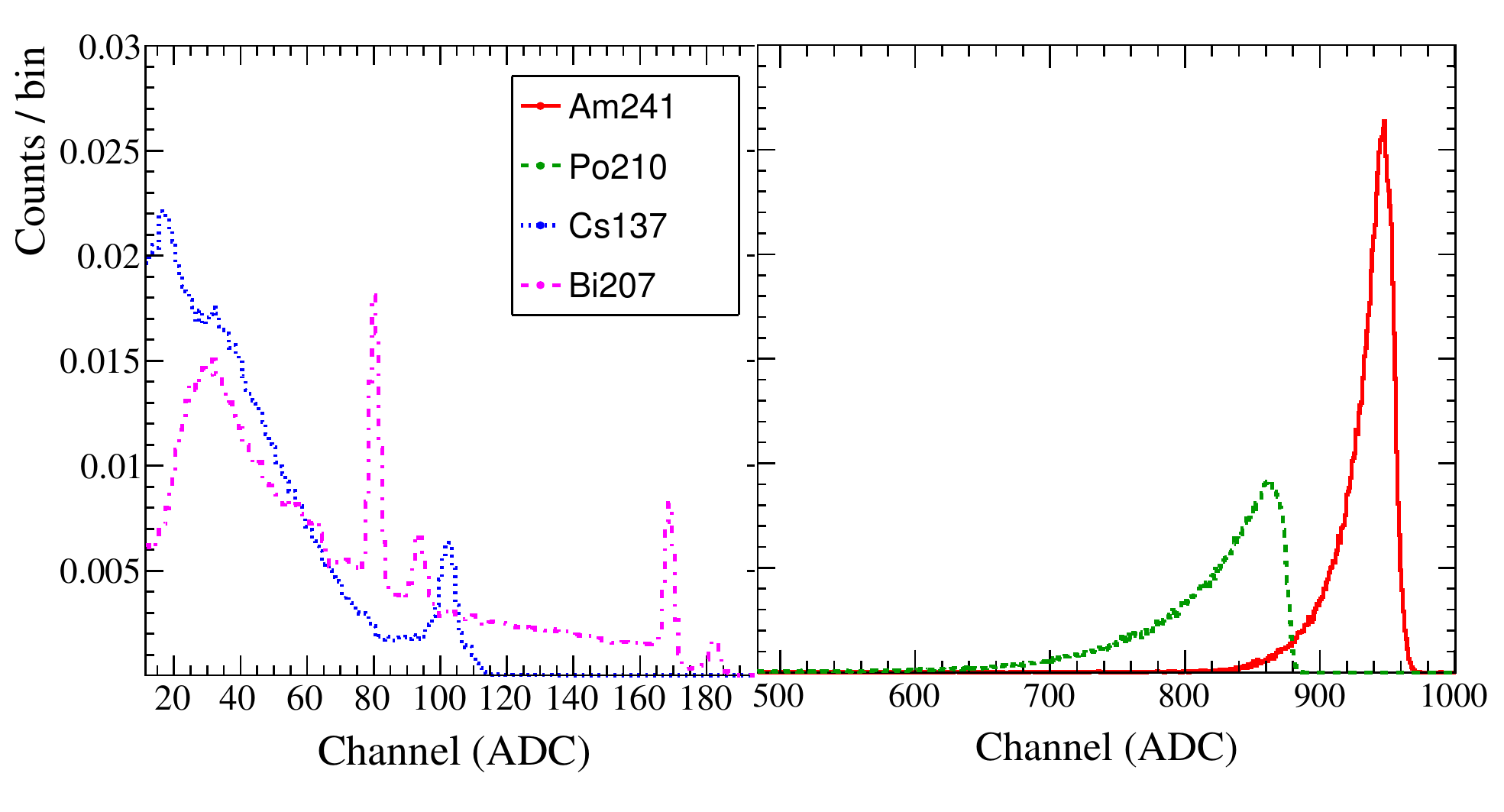}
\caption{\label{fig:ecaldata} Data used for the energy calibration.
  The curves on the left show the IC electron source data, and the
  curves on the right show the $\alpha$ source data.}
\end{center}
\end{figure*}

A $^{241}$Am surface $\alpha$ source, a $^{210}$Po thin window
$\alpha$ source, a thin window $^{137}$Cs electron conversion source,
and a thin window $^{207}$Bi electron conversion source with known
particle energies were used to calibrate the energy scale of the Si
surface barrier detector.  The calibration sources were placed close
to the Si instead of the $^6$Li glass in the setup.  The two electron
conversion sources had more than one energy possible, resulting in
more than one energy peak.  Area normalized MCA output from the four
energy calibration sources are shown in Fig.~\ref{fig:ecaldata}. Each
of the peaks was fitted individually, and compared to the known
energies of the source, as summarized in Table~\ref{tab:ecaldata}.  

\begin{table}[hb]
\begin{center}
\begin{tabular}{ | l | c | c | }
  \hline
  Peak Source  &  $E_{true}$ (keV) & Mean Fit ADC \\ 
  \hline
$^{207}$Bi 570~keV $K_{ab}$ IC  & 479.181 &  80.3 $\pm$ 2.0 \\
$^{207}$Bi 570-keV $L_{1ab}$ IC & 553.309 &  93.7 $\pm$ 2.1 \\
$^{137}$Cs $K_{ab}$ IC          & 625.698 & 102.0 $\pm$ 2.0 \\
$^{207}$Bi 1064~keV $K_{ab}$ IC & 973.141 & 168.7 $\pm$ 2.0 \\
$^{207}$Bi 1064~keV $L_{1ab}$ IC& 1047.27 & 182.1 $\pm$ 2.1 \\ 
$^{210}$Po $\alpha$            & 4965    & 860.1 $\pm$ 4.7 \\
$^{241}$Am $\alpha$            & 5486    & 947.7 $\pm$ 2.1 \\
  \hline
\end{tabular}
\end{center}
\caption{Peak fit results used for the rough energy calibration.  For
  the Am source, the energy of the dominant decay was used, and for
  the Po source, whose $\alpha$ nominally has an energy of 5307~keV, a
  value of 4965~keV was used to account for a larger air gap and the
  presence of a film of mylar film on the source. }\label{tab:ecaldata}
\end{table}

It is well known that scintillators exhibit ionization quenching for
large energy deposit (characterized by Birks' constant), as is the
case for the alpha and triton produced in neutron captures in our
lithium glasses \cite{AWDalton}.  This is not a concern in our
energy calibration, since we are not looking at the scintillation
light in these tests.  Instead, we are calibrating the response of a
surface barrier detector to different ionization energy deposits.
Rather than measuring the scintillation light, we are measuring the
energy spectrum of alpha or triton particles that escape from the GS20
side of the glass.

Note that the Po source was assigned a larger uncertainty for two
reasons. One was the alignment of the source and detector were not
well controlled.  The active source area was small and difficult to
line up with the Si portion of the detector, which could lead to some
energy loss in the Po energy spectrum.  Another reason was that the
source has a mylar film leading to a larger energy loss for this
source.

\begin{figure}[htpb]
\begin{center}
\includegraphics[width=0.49\textwidth]{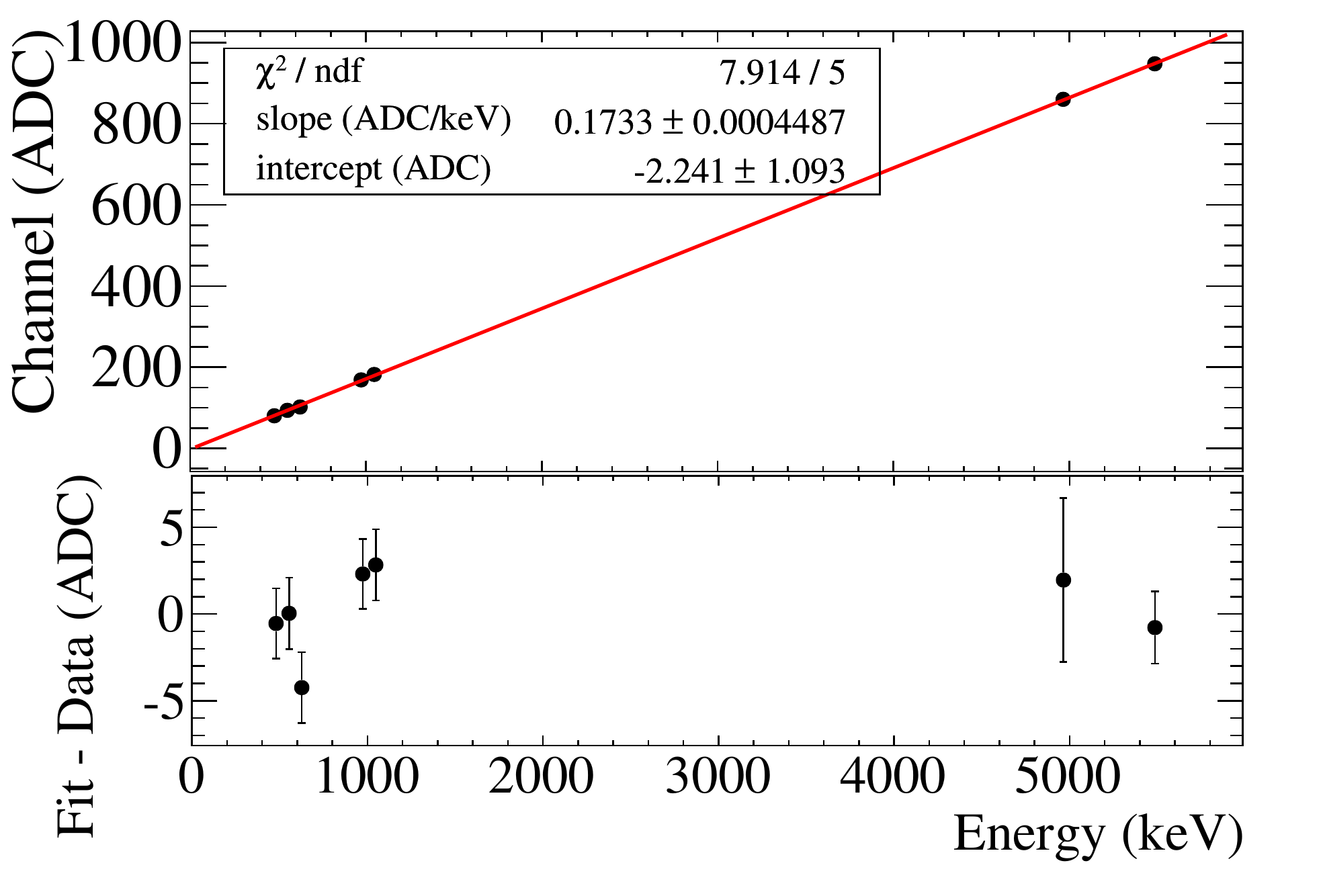}
\caption{\label{fig:ecal} The channel versus energy calibration plot
  is in the top plot, and the bottom plot shows the residual from the
  linear fit.}
\end{center}
\end{figure}

\section{Results}

\subsection{Background}

Since the rate of thermal neutron capture is low by design, background
had a larger presence in the neutron data than in the calibration
data.  Background events come from neutron captures on materials
around the detector, x-rays from the neutron source, other radiation
from neutron captures on other materials near the detector, natural
radioactivity in the vicinity of the detector, cosmic rays, and
electronic noise.

Removing the glass and the sources yielded some background, but adding
the neutron source back into the set-up without the lithium glass
showed that the largest background was due to the presence of the
source.  These could be neutron captures on materials around the
detector and from x-rays from the neutron sources.  In any case the
background can be measured and used as a reference when comparing to
data collected with the lithium glass added.

\subsection{Signal}

\begin{figure}[htpb]
\begin{center}
\includegraphics[width=0.49\textwidth]{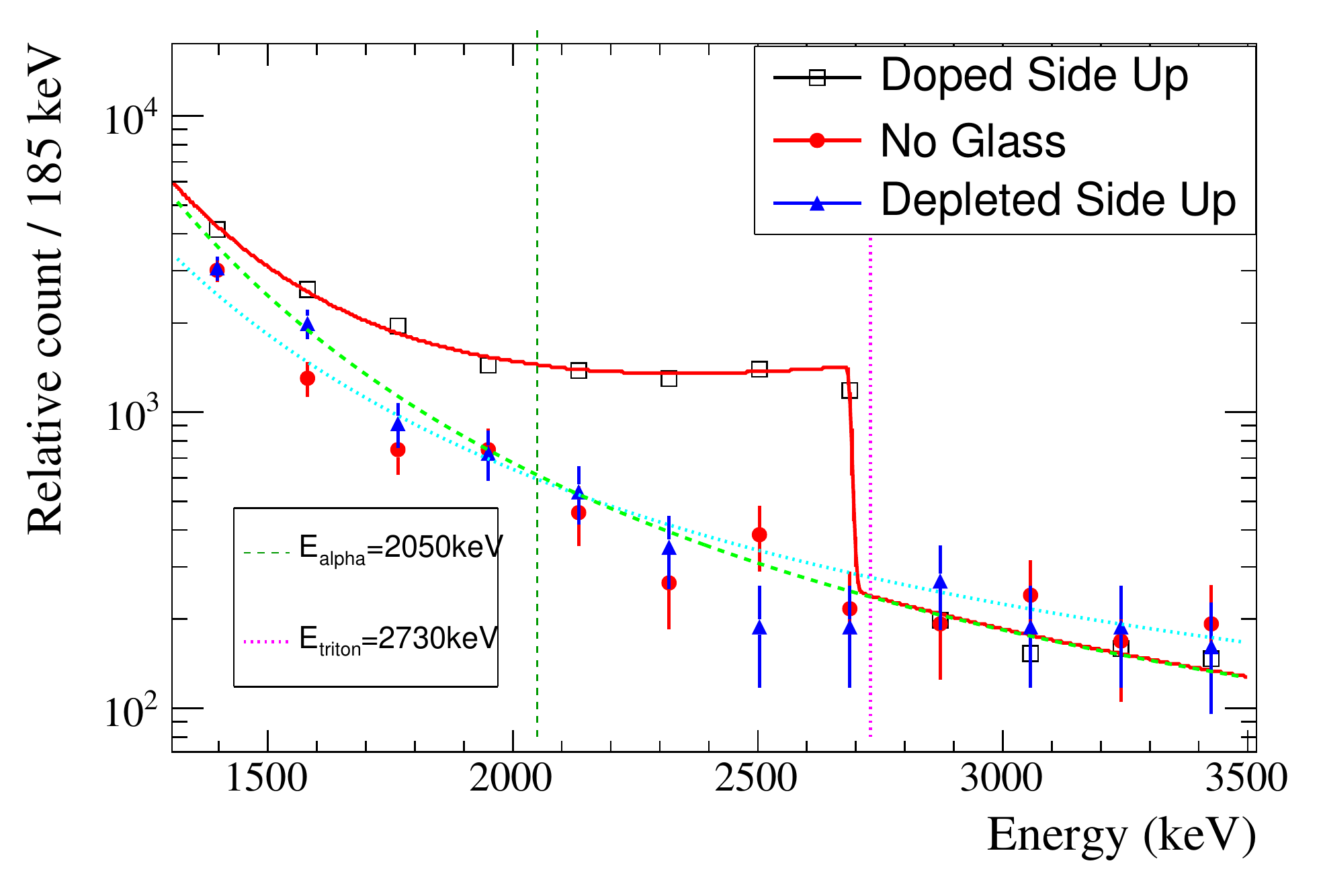}
\caption{\label{fig:lithiumside} Energy deposit in the Si detector for
  a background run (red circles), a run with the $^{6}$Li depleted
  side facing the Si detector (blue triangles), and the sum of all
  runs with the $^{6}$Li enriched side facing the Si detector (black
  squares).  The solid curved line shows the best fit to the neutron
  capture data, the dotted curved line shows the fit to the depleted
  side data, and the curved dashed line shows the background after the
  fit using the neutron capture data.}
\end{center}
\end{figure}

The lithium glass stack was placed in the set-up with either side up.
Data was collected for 24 hours for a particular side of the glass.
The resulting spectra were then normalized and compared to background.
Area normalized energy spectra for the background run, and different
orientations of the $^{6}$Li glass are shown in
FIG.~\ref{fig:lithiumside}.

While the $^6$Li depleted side of the glass was similar to the
background, the $^6$Li doped side showed an excess of events between
1300~keV and 2900~keV.  This energy range is higher than the $\alpha$
energy, indicating that most of the excess detected is due to triton
captures.  This makes sense, as the range of the triton is longer, and
so contributes more to the detected spectrum.  This procedure was
repeated with nine other pieces of glass.  Each time an excess
appeared around the same energy range for one side of the glass,
allowing us to confirm which side of the glass contained the doped
scintillator.

A fit was performed in order to estimate the $^{6}$Li neutron capture
signal, modelled as a sum of an alpha and a triton as a pair of
gaussians with exponential tails, to a background modelled as an
exponential of the form $e^{A/E+B}$, with $A$ and $B$ determined from
a fit to the depleted side facing the detector.  First a fit to the
sum of all of neutron captures from ten samples that were tested was
performed, where all of the parameters in the fit were allowed to
vary.  When fitting each sample individually, all of the parameters
were fixed to the values found from the fit to all of the samples,
except the number of background, the number of alpha and number of
triton which were allowed to float.  The different number of signal
and background events for each sample are due to the different
counting times used.  To account for uncertainties in the model, the
uncertainties in the number of counts determined by integrating the
function over the range 1.3~MeV to 2.9~MeV was increased by a factor
$\sqrt{\chi^2/\textrm{DOF}}$ determined by the ratio of the fit
$\chi^2$ to degrees of freedom (DOF) .

A summary of the fit results shown in Table~\ref{tab:fitresults} shows
that the neutron capture signal is always measured, but that there is
a fair bit of variation due to the number of uncontrolled variables
such as the inexact placement of glass in air near the detector, and
the change in placement of wax blocks around neutron source.

\begin{table*}[htb]
\begin{center}
\begin{tabular}{  c | c  c  | c  c  c  c | }
  \hline
Sample   & $\chi^2$ / DOF  &  p-value &   Signal ($S$)  &  Background ($B$) &  $S / B$\\ 
  \hline
All      &  $20.50 /    6$  &  0.002     &  $5527 \pm 235$ &  $5299 \pm 135$   & $1.043 \pm  0.044$\\
  \hline
 0       & $17.26 /    9$   &  0.045     &  $1206 \pm  75$ &  $867 \pm  41$    & $1.39 \pm  0.09$\\
 1       & $16.73 /    9$   &  0.053     &   $402 \pm  41$ &  $257 \pm  22$    & $1.57 \pm  0.16$\\
 2       & $11.21 /    9$   &  0.262     &   $526 \pm  37$ &  $293 \pm  19$    & $1.80 \pm  0.13$\\
 3       & $15.01 /    9$   &  0.091     &   $315 \pm  41$ &  $355 \pm  24$    & $0.89 \pm  0.12$\\
 4       &  $9.90 /    9$   &  0.359     &   $878 \pm  50$ &  $707 \pm  28$    & $1.24 \pm  0.07$\\
 5       &  $4.78 /    9$   &  0.853     &  $1039 \pm  40$ &  $970 \pm  23$    & $1.07 \pm  0.04$\\
 6       &  $8.40 /    9$   &  0.494     &   $422 \pm  31$ &  $316 \pm  17$    & $1.34 \pm  0.10$\\
 7       &  $2.82 /    9$   &  0.971     &   $413 \pm  18$ &  $310 \pm  10$    & $1.33 \pm  0.06$\\
 8       &  $9.59 /    9$   &  0.385     &   $444 \pm  38$ &  $449 \pm  22$    & $0.99 \pm  0.08$\\
 9       &  $6.97 /    9$   &  0.640     &   $347 \pm  28$ &  $315 \pm  16$    & $1.10 \pm  0.09$\\
  \hline
\end{tabular}
\end{center}
\caption{Results from fitting for the total number of alpha and triton
  particles above the background for each of the ten samples tested,
  and for the sum of all ten samples.  Refer to the text for
  details.}\label{tab:fitresults}
\end{table*}

\section{Conclusion}

Using a Si surface barrier detector and a thermal neutron source, it is
possible to determine the presence of $^6$Li in scintillating lithium
glass.  Neutrons capture on the $^6$Li to produce an $\alpha$ and
triton.  When the neutron captures near the surface of the glass, the
ejected ions can be detected over background, and thus determine which
side of the glass to put facing toward the UCN source when assembling
the detector.

\section{Acknowledgements}

We would like to acknowledge the Natural Sciences and Engineering
Research Council of Canada for funding.  We thank the TRIUMF-UCN group
for useful discussions relating to the neutron Electric Dipole Moment
experiment being planned, and discussions on the requirements for the
UCN detector.  We also would like to acknowledge Chuck Davis' idea of
using an ion chamber to check the lithium glass side by trying to
detect the products from the neutron captures.  We have adapted his
idea to the experiment described in this paper.





\bibliographystyle{elsarticle-num}
\bibliography{<your-bib-database>}



\end{document}